\documentstyle[eqsecnum,preprint,aps,epsbox]{revtex}

\begin{document}

\draft
\preprint{YITP-98-63, WU-AP/74/98, gr-qc/9811024}
\title{Monopole Inflation in Brans-Dicke Theory}
\author{Nobuyuki Sakai\thanks{Electronic address: 
sakai@yukawa.kyoto-u.ac.jp}, Jun'ichi Yokoyama}
\address{Yukawa Institute for Theoretical Physics, Kyoto University, Kyoto
606-8502, Japan}
\author{and}
\author{Kei-ichi Maeda}
\address{Department of Physics, Waseda University, Tokyo 169-8555, Japan}
\date{Revised 25 January 1998}
\maketitle

\begin{abstract}

According to previous work, topological defects expand exponentially without 
an end if the vacuum expectation value of the Higgs field is of the order 
of the Planck mass. We extend the study of inflating topological defects to the 
Brans-Dicke gravity. With the help of numerical simulation we investigate 
the dynamics and spacetime structure of a global monopole. Contrary to 
the case of the Einstein gravity, any inflating monopole eventually shrinks 
and takes a stable configuration. We also discuss cosmological constraints 
on the model parameters.

\end{abstract}

\vskip 1cm
\begin{center}
To appear in {\it Physical Review D}\\
PACS number(s): 04.50.+h, 98.80.Cq
\end{center}

\newpage
\tighten
\section{Introduction}

For the last decade spacetime solutions of 
gravitating  monopoles has been intensively 
studied in the literature \cite{BV,HL,mm,GR,Lin,Vil,SSTM,Sak,CV}. 
This originated from rather 
mathematical interest in static monopole solutions. In both cases of 
global monopoles \cite{BV,HL} and of magnetic monopoles \cite{mm}, static 
regular solutions are nonexistent if the vacuum expectation value
(VEV) of 
the Higgs field $\eta$ is larger than a critical value $\eta_{{\rm 
sta}}$, which is of the order of the Planck mass, $m_{{\rm Pl}}$. The 
properties of monopoles for $\eta>\eta_{{\rm sta}}$ had been a puzzle 
for some time.

In connection with the above issue, it has been claimed by several authors
independently that monopoles expand exponentially if 
$\eta>O(m_{{\rm Pl}})$ \cite{GR,Lin,Vil}.  Among them, Guendelman 
and Rabinowitz \cite{GR} assumed the simplified model 
of a global monopole under
the thin-wall approximation, 
while Linde \cite{Lin} and Vilenkin \cite{Vil} made qualitative 
arguments on any kinds of topological defects and argued that their core
would inflate if the VEV of the Higgs field is large. 
Later the full evolution equations 
for monopoles were numerically solved \cite{SSTM,Sak}, supporting the 
previous arguments as a whole. Furthermore, those numerical analyses 
revealed an unexpected result that the critical value, $\eta_{{\rm
inf}}$, above which inflation occurs in the core, is larger than
$\eta_{{\rm sta}}$, and stable but nonstatic solutions exist for
$\eta_{{\rm sta}}<\eta<\eta_{{\rm inf}}$ \cite{Sak}. 
For example, in the case of
global monopoles,  we find
$\eta_{{\rm sta}}\cong 0.20m_{{\rm Pl}}$ and 
$\eta_{{\rm inf}}\cong 0.33m_{{\rm Pl}}$ \cite{Sak}. 
Global spacetime structure of 
an inflating monopole has also been discussed in \cite{SSTM,Sak,CV}.

In this paper, we extend the study of monopole inflation to the Brans-Dicke
(BD) theory. If the BD field exists, the inflationary universe may not 
expand exponentially but expand with a power law even in the presence of 
an effective cosmological constant \cite{LS,BM}. For example, in 
extended inflation \cite{LS}, this slower
expansion was expected to solve the graceful 
exit problem of old inflation \cite{oi}. We thus expect 
that the BD field also affects the dynamics and global spacetime structure 
of inflating monopoles. 

We would also like to discuss whether the present model can be a 
realistic cosmological model or not. Once inflation happens, the 
spatial gradients of the fields become negligibly small on the scale of the 
observed universe. Therefore, just as other inflationary models, we 
have only to care about quantum fluctuations. This issue resolves itself 
into the commonplace analysis of density perturbations in the 
inflationary universe. Here we adopt the result obtained by Starobinsky 
and Yokoyama \cite{SY}, who derived the general expression of the 
density fluctuations in the BD theory, 
taking account of the isocurvature mode as well as the adiabatic mode. 

The rest of the paper is composed of two independent analyses. In the first 
part (\S2), we investigate the dynamics and global spacetime structure of 
an inflating global monopole. In the second part (\S3), we discuss 
cosmological constraints on this model, which come mainly from 
density perturbations. We use the units $c=\hbar=1$ throughout the paper.

\section{Dynamics and Spacetime Structure of a Global Monopole}

The Brans-Dicke-Higgs system, which we consider here, is described by the action
\begin{equation}\label{action}
  S=\int d^4 x \sqrt{-g} \left[\frac{\Phi}{16\pi}{\cal R}
     -\frac{\omega}{16\pi\Phi}(\nabla_{\mu}\Phi)^2
     -\frac12(\nabla_{\mu}\Psi^a)^2-V(\Psi)\right],
\end{equation}
with
\begin{equation}\label{pote}
V(\Psi)= {\lambda\over 4}(\Psi^2-\eta^2)^2, ~~ 
\Psi\equiv\sqrt{\Psi^a\Psi^a},
\end{equation}
where $\Phi$ and $\Psi^a$ are the BD field and the real triplet Higgs 
field, respectively. $\omega$ and $\lambda$ are the BD parameter and the 
Higgs self-coupling constant, respectively.

Let us begin with a discussion of the fate of an inflating topological defect. 
Once inflation begins, the core region can be approximated by 
the flat Friedmann-Robertson-Walker
spacetime:
\begin{equation}\label{FRW}
ds^2=-dt^2+a(t)^2d{\bf x}^2,
\end{equation}
which yields the solution of extended inflation \cite{MJ,LS}: 
\begin{equation}\label{extinf}
a(t)\propto\left(1+{H_i t\over\alpha}\right)^{\omega+\frac12}, ~~~
\Phi(t)\propto\left(1+{H_i t\over\alpha}\right)^2,
\end{equation}
where $H_i\equiv\sqrt{8\pi V(0)/3\Phi_i}$ is the Hubble parameter at the 
beginning of inflation, and $\alpha^2\equiv(2\omega+3)(6\omega+5)/12$. 
Hence, the effective Planck mass,
\begin{equation}
m_{\rm Pl}(\Phi)\equiv\sqrt{\Phi}\propto a^{{1\over\omega+\frac 12}}, 
\end{equation}
continues to increase until inflation ends. This implies that, even if 
$\eta/m_{\rm Pl}(\Phi)$ is large enough to start inflation initially, 
it eventually becomes smaller than the critical value 
$(\cong 0.33)$ \cite{SSTM}. 
We thus speculate that any defect eventually shrinks after inflation. 

In order to ascertain the above argument, we carry out numerical analysis 
for spherical global monopoles. The coordinate system we adopt is
\begin{equation}\label{metric}
ds^2=-dt^2+A^2(t,r)dr^2+B^2(t,r)r^2(d\theta^2+\sin^2\theta d\varphi^2).
\end{equation}
For the matter fields, we adopt the hedgehog ansatz:
\begin{equation}\label{hg}
\Psi^a=\Psi(t,r)\hat r^a,~~~
\hat r^a \equiv (\sin\theta\cos\varphi,\sin\theta\sin\varphi,\cos\theta).
\end{equation}
For the initial configurations of the Higgs fields and the BD field, we assume
\begin{equation}\label{Psi0}
\Psi(t=0,r)=\eta\tanh\Bigl({r\over\delta}\Bigr) ~~ {\rm with} ~~
\delta\equiv{\sqrt{2}\over\sqrt{\lambda}\eta}, ~~~
\dot\Psi(t=0,r)=0,
\end{equation}
\begin{equation}\label{Phi0}
\Phi(t=0,r)=\Phi_i={\rm const.}, ~~~~ \dot\Phi(t=0,r)=0,
\end{equation}
where an overdot denotes $\partial/\partial t$. 

Although the numerical code developed by one of us \cite{Sak} worked well for 
monopoles in the Einstein theory, for the present system which 
includes the BD field it sometimes does not keep good accuracy for 
sufficient time. We therefore improve our numerical method, mainly based 
on the idea of Nakamura {\it et al.}\cite{Nak}~ The basic
equations and our improved numerical method are presented in Appendix.

To illustrate our results, we take $\omega=1,~ \lambda=0.1$ and 
$\eta/m_{\rm Pl}(\Phi_i)=1$. The evolutions of $\Psi$ and 
$\eta/m_{\rm Pl}(\Phi)$ are shown in Figure 1. Clearly the above arguments 
are verified by this numerical result: as $\eta/m_{\rm Pl}(\Phi)$ gets close to 
the critical value, the monopole stops expanding and turns to shrink. 
In Fig. 2 we plot the trajectories of the position of $\Psi=\eta/2$ 
for several values of $\omega$. The curves indicate that a monopole do 
not shrink toward the origin but tends to take a stable configuration.

Next, we shall examine the metric outside the monopole. Before we 
consider monopoles in the BD theory, let us review previous work on the 
outer solution in the Einstein theory briefly. The approximate expression 
for the metric outside a static monopole was found by Barriola and Vilenkin 
(BV) \cite{BV}:
\begin{equation}\label{BV}
ds^2=-\left(1-\Delta-{2M_c\over m_{\rm Pl}^{~2} R}\right)dT^2
+\left(1-\Delta-{2M_c\over m_{\rm Pl}^{~2} R}\right)^{-1}dR^2
+R^2(d\theta^2+\sin^2\theta d\varphi^2).
\end{equation}
where $M_c$ is an integration constant and 
$\Delta\equiv 8\pi\eta^2/m_{{\rm Pl}}^{~2}$. 
Later Harari and Lousto pointed out that the mass parameter $M_c$ is
negative and therefore a global monopole has a repulsive nature \cite{HL}. 
For large $R$, the metric (\ref{BV}) can be approximated by
\begin{equation}\label{BV2}
ds^2=-\left(1-\Delta\right)dT^2+\left(1-\Delta\right)^{-1}dR^2
+R^2(d\theta^2+\sin^2\theta d\varphi^2),
\end{equation}
and therefore $4\pi\Delta$ is interpreted as a deficit solid angle. 
Although the expression (\ref{BV2}) looks singular when
$\Delta=1$, Cho \& Vilenkin \cite{CV} showed that it is just an
apparent singularity and (\ref{BV2}) also applies to the case $\Delta>1$. 

Now we would like to discuss the properties of the outer spacetime by 
comparing it with the BV solution (\ref{BV}). As a 
coordinate-independent variable, we adopt the Misner-Sharp mass 
\cite{MS}, which is defined as
\begin{equation}\label{MS}
\tilde M\equiv{\bar R\over 2}(1-g^{\mu\nu}\bar R_{,\mu}\bar R_{,\nu}),
\end{equation}
where $\bar R\equiv\sqrt{g_{\theta\theta}}$ is the circumferential
radius of the spacetime. Note that $\tilde M$ has a dimension of length, and 
$m_{\rm Pl}^{~2}\tilde M$ is a mass in a usual sense. The BV solution 
is characterized by $\tilde M=\tilde M_c+\Delta R$ with 
$\tilde M_c=M_c/m_{\rm Pl}^{~2}$.

Figure 3 shows a plot of $\tilde M/\bar R$. In the case of the Einstein 
theory (a), $\tilde M/\bar R$ seems to converge into $\Delta/2$ 
(=$4\pi$ in this case) at large $R$, confirming that the metric is well 
approximated by (\ref{BV}). The result with the BD gravity looks
similar, but a different character is that $\tilde M/\bar R$ decreases 
with time even at a far region. This behavior is consistent with the 
BV solution in the Einstein theory: if we take account of the 
time-variation of $m_{\rm Pl}$ in the BV expression (\ref{BV}), a deficit solid 
angle effectively decreases and (\ref{BV}) also approximates the solution 
in the BD theory.

Our results obtained so far are rephrased as follows: our numerical 
integration of the evolution equations verifies as a whole the arguments 
with considering only the time-variation of $m_{\rm Pl}(\Phi)$. That is, 
the metric in the core of a monopole is well approximated by 
(\ref{extinf}), and the metric outside the core by (\ref{BV}) with 
changing $m_{\rm Pl}$. Although we have chosen small $\omega$ to illustrate 
the effect of the BD field clearly, if we take $\omega>500$ as constrained 
by observation \cite{Rea}, the spatial gradient of $\Phi$ is much smaller, 
justifying more the arguments with neglecting the spatial configuration of 
$\Phi$.

The result that inflation eventually ends stems merely from the change 
of the local values of $m_{\rm Pl}(\Phi)$. We may therefore extend this 
result to models of other topological defects.

\section{Cosmological Constraints}

In this section we discuss constraints for this model to be a realistic 
cosmological model. As we mentioned in \S1, the main 
constraint is from  density perturbations. In the 
present model matter fields are composed of the BD field $\Phi$ and 
the Higgs field $\Psi^a$. For the Higgs field, we only calculate 
fluctuations in the radial direction $\delta\Psi$, because
fluctuations along the other directions do not contribute to growing
modes.  

Here we analyze the field equations in the Einstein frame by use of 
a conformal transformation, following previous work \cite{BM,SY}. 
A wide class of generalized Einstein theories is described 
by the Einstein-Hilbert action with scalar fields:
\begin{equation}\label{action2}
\hat{\cal S} = \int d^4 \hat x \sqrt{-\hat g} \biggl[{\hat{\cal 
R}\over2\kappa^2} -{1\over2}(\hat\nabla\phi)^2
-{e^{-\gamma\kappa\phi}\over2}(\hat\nabla\Psi)^2 
- e^{-\beta\kappa\phi}V(\Psi)-U(\phi) \biggr]
~~ {\rm with} ~~ \kappa^2\equiv8\pi G,
\end{equation}
where $\beta$ and $\gamma$ are constants, and $\phi$ is the 
redefined scalar field.

In the BD theory, on which we concentrate in this section, the original 
action in the Jordan
frame (\ref{action}) is transformed into (\ref{action2}) with
\begin{equation}
\gamma={\beta\over 2}=\sqrt{{2\over2\omega+3}},~~~ U(\phi)=0
\end{equation}
through the conformal transformation and the redefinition of the scalar field
\begin{equation}\label{contra}
\hat g_{\mu\nu} = {\Phi\over\Phi_0} g_{\mu\nu},~~~
\sqrt{\gamma}\kappa\phi \equiv \ln{\Phi\over\Phi_0},
\end{equation}
with $\Phi_0$ being a constant.
From (\ref{contra}), the Planck mass in the Jordan frame is written as
\begin{equation}
m_{\rm Pl}(\Phi)\equiv\sqrt{\Phi}=\sqrt{\Phi_0}e^{\gamma\kappa\phi/2}.
\end{equation}
We set $\phi=0$ at the present time, $t=t_0$, identifying a constant
$\sqrt{\Phi_0}$ with the Planck mass at present. Because the 
evolution of the BD 
field after inflation is negligibly small for $\gamma^2\ll 1$, 
we may identify $\Phi$ at the end of inflation, $\Phi_f$, with 
$\Phi_0$ ({\it i.e.}, $\phi_f\cong\phi_0=0$). In the following we sometimes 
use the symbol $m_{\rm Pl}(\Phi)$ instead of $\phi$, and define 
$m_{\rm Pl,0}\equiv m_{\rm Pl}(\Phi_0)\cong m_{\rm Pl}(\Phi_f)$.

Hereafter, we omit a hat, which has denoted quantities in the Einstein frame. 
Taking the background as the spatially flat Friedmann-Robertson-Walker
spacetime (\ref{FRW}), the field equations for the homogeneous parts read
\begin{equation}
H^2\equiv\left({\dot a\over a}\right)^2={\kappa^2\over 3}
\left(\frac12 e^{-\gamma\kappa\phi}\dot\Psi^2+\frac12\dot\phi^2
+e^{-2\gamma\kappa\phi} V\right),
\end{equation}
\begin{equation}
\ddot\Psi+3H\dot\Psi-2\gamma^2\dot\phi\dot\Psi
+e^{-\gamma\kappa\phi}{dV\over d\Psi}=0,
\end{equation}
\begin{equation}
\ddot\phi+3H\dot\phi+{\gamma\kappa\over 2}e^{-\gamma\kappa\phi}\dot\Psi^2
-2\gamma\kappa e^{-2\gamma\kappa\phi} V=0.
\end{equation}
The conditions for slow-roll inflation ($|\ddot\Psi|\ll|3H\dot\Psi|$,
$|\ddot\phi|\ll|3H\dot\phi|$, etc.) are equivalent to
\begin{equation}\label{slow1}
\gamma^2\ll\frac 32, ~~ 
\left|{m_{\rm Pl}(\Phi) V_{,\Psi}\over V}\right|\ll\sqrt{48\pi},
~~
\left|{m_{\rm Pl}(\Phi)^2V_{,\Psi\Psi}\over V}\right|\ll 24\pi,
\end{equation}
where $V_{,\Psi}\equiv dV/d\Psi$.
For the potential (\ref{pote}), the second and the third 
inequalities in (\ref{slow1}) in the range $0<\Psi<\eta$ are
equivalent to
\begin{equation}\label{slow2}
\epsilon^{-1}\equiv{\sqrt{6\pi}\eta\over m_{\rm Pl}(\Phi)}\gg 1,
~~
\Psi\ll\eta\left(1-{\epsilon\over 2}\right)\equiv\Psi_f,
\end{equation}
where terms of higher-order of $\epsilon$ have been neglected in the second 
inequality.

One of the distinguished features of this model is that there are two 
scenarios of exiting from an inflationary phase and entering a 
reheating phase. This is illustrated by the slow-roll conditions (\ref{slow2}). 
When the condition $\epsilon^{-1}\gg1$ (a more precise condition is 
$\eta/m_{\rm Pl}(\Phi)>0.33$) breaks down, inflation stops globally. Before this 
time, many local regions with the present-horizon size enters a 
reheating phase when the second condition $\Psi\ll\Psi_f$ breaks down. The 
second scenario is just like that of standard topological inflation or other 
slow-roll inflationary models. In the first scenario a microscopic monopole 
might remain in the observable universe. Unfortunately, however, this 
possibility turns out to be ruled out because the spectrum of density 
fluctuation is tilted excessively if $\eta$ is too close to the 
critical value $0.33m_{\rm Pl}$ \cite{EKOY} and because
$\eta/m_{\rm Pl}(\Phi_f)$ must be larger than 1 from the COBE normalization as
will be seen later in Fig. 4. Thus we only consider the second (standard) 
reheating scenario below.

When inequalities (\ref{slow2}) are satisfied, the field equations are 
approximated by
\begin{equation}\label{sloweq1}
H^2={\kappa^2\over 3}e^{-2\gamma\kappa\phi} V,
\end{equation}
\begin{equation}\label{sloweq2}
3H\dot\Psi=-e^{-\gamma\kappa\phi}{dV\over d\Psi},
\end{equation}
\begin{equation}\label{sloweq3}
3H\dot\phi=2\gamma\kappa e^{-2\gamma\kappa\phi} V,
\end{equation}
which yield solutions \cite{BM,SY},
\begin{equation}\label{con1}
N\equiv\ln\Bigl({a_f\over a}\Bigr)=-{\kappa\phi\over2\gamma},
\end{equation}
\begin{equation}\label{con2}
N\cong{1-e^{-2\gamma^2 N}\over 2\gamma^2}
=\kappa^2\int^{\Psi}_{\Psi_f}{V\over V_{,\Psi}}d\Psi
={\pi\over m_{\rm Pl,0}^{~2}}
\left(2\eta^2\ln{\Psi_f\over\Psi}+\Psi^2-\Psi_f^2\right).
\end{equation}

Starobinsky and Yokoyama \cite{SY} derived the amplitude of density perturbation 
on comoving scale $l=2\pi/k$ in terms of Bardeen's variable $\Phi_A$ \cite{Bar} 
as
\begin{equation}\label{SY}
\Phi_A(l)^2={48 V\over 25m_{\rm Pl}(\Phi)^4}
\left[{8\pi V^2\over m_{\rm Pl}(\Phi)^2 V_{,\Psi}^{~2}}
+{(e^{2\gamma^2 N}-1)^2\over 4\gamma^2}\right],
\end{equation}
where all quantities are defined at the time $t_k$ when $k$-mode leaves the 
Hubble horizon, {\it i.e.}, when $k=aH$. Since the large-angular-scale 
anisotropy of the microwave background 
due to the Sachs-Wolfe effect is given by $\delta T/T=\Phi_A/3$, we 
can constrain the values of $\lambda$ and $\eta/m_{\rm Pl}$ by the
COBE-DMR data normalization \cite{Ben},
\begin{equation}\label{Q}
\frac{1}{3}\Phi_A\cong 10^{-5},
\end{equation}
on the relevant scale.
To calculate (\ref{Q}) explicitly, we have to relate scales during inflation and 
present scales. The number of $e$-folds $N$ between $t=t_k$ and $t=t_f$ is given 
by
\begin{equation}\label{Nk}
N_k=(1-\gamma^2)^{-1}
\left(68-\ln{k\over a_0H_0}+\ln{V_f^{\frac 14}\over m_{\rm Pl,0}}
+2\ln{V_k^{\frac 14}\over V_f^{\frac 14}}
-\frac 13\ln{V_f^{\frac 14}\over\rho_{{\rm rh}}^{\frac 14}}\right),
\end{equation}
where $\rho_{{\rm rh}}$ is the energy density when 
reheating completes, which is not determined without specifying a reheating 
model. Here we choose $N_{k=a_0H_0}=65$ typically. Then the 
corresponding values of $\phi$ and $\Psi$ are determined from 
(\ref{con1}) and (\ref{con2}). Figure 4 shows the allowed values of 
$\lambda$ and $\eta/m_{\rm Pl}$ for $\gamma=0.045$ ($\omega=500$). The 
concordant values are represented by two curves, which
is a distinguished feature for the double-well potential (\ref{pote}). 
Unfortunately, fine-tuning of $\lambda ~(
~\mbox{\raisebox{-1.0ex}{$\stackrel{\textstyle <}{\textstyle \sim}$ }}
10^{-13})$ is needed, just like other models \cite{SY}. The condition of 
$\lambda
~\mbox{\raisebox{-1.0ex}{$\stackrel{\textstyle <}{\textstyle \sim}$ }}
10^{-13}$ in 
the present potential was also found in \cite{GL}. 
The constraints for $\omega>500$ are practically no 
different from those in the Einstein gravity. 

Using the relation $d\ln k=da/a+dH/H$ and (\ref{sloweq1})-(\ref{sloweq3}), we 
obtain
\begin{equation}\label{index}
n-1\equiv{d\ln\Phi_A^2\over d\ln k}
\cong-{3m_{\rm Pl}(\Phi)^2 V_{,\Psi}^{~2}\over 8\pi V^2}
+{m_{\rm Pl}(\Phi)^2 V_{,\Psi\Psi}\over 4\pi V}
-6\gamma^2.
\end{equation}
The spectral indices are plotted in Fig. 5; the two lines correspond 
to the two lines of COBE-normalized amplitudes in Fig. 4. As 
(\ref{index}) suggests, the deviation from the Einstein theory is only 
$6\gamma^2\cong 0.01$ for $\omega=500$. Therefore, relatively large shift 
from $n=1$ is caused not by the BD field but by the double well potential 
(\ref{pote}). 

Besides the amplitude of density perturbation, we have to check other conditions 
for successful inflation. First, 
if we take quantum fluctuations $\delta\Psi_{{\rm Q}}$ into account, 
the classical dynamics is meaningless unless
\begin{equation}\label{psiq}
{|\dot\Psi|\over H}>\delta\Psi_{\rm Q}={e^{\gamma\kappa\phi/2}H\over 2\pi}.
\end{equation}
We check the value of $\dot\Psi$ at $k=a_0H_0$ which satisfies the COBE
normalization, finding that (\ref{psiq}) is always satisfied.

The second is the condition that classical description of spacetime is valid. We 
examine this because we have considered the case of $\eta\cong m_{\rm Pl}$, 
which may be critical. We require
\begin{equation}\label{cg}
V(\Psi_i)<m_{\rm Pl}(\Phi_i)^4.
\end{equation}
This condition is also satisfied for $\eta\cong m_{\rm Pl,0}$ and $\lambda 
~\mbox{\raisebox{-1.0ex}{$\stackrel{\textstyle <}{\textstyle \sim}$ }}
10^{-13}$. In the end the two conditions do not add new constraints on the 
model.

Now, let us discuss the detectability of relic monopoles in this 
model. First, we investigate how the long-range term of the monopole mass
density, $\rho_{\rm gm}(R)=\eta^2/R^2$, contributes to the cosmological 
mass density, where $R$ is the distance from a monopole core to an 
observer. Hiscock \cite{His} derived the condition that the averaged mass 
density of multiple global monopoles does not exceed the cosmological mass 
density, {\it i.e.},
$\langle\rho_{\rm gm}\rangle<\rho_0$. His result is equivalent to
\begin{equation}
{N\over V_H}<5\times 10^{-4}\Omega_0^{\frac32}\left({m_{\rm Pl,0}\over\eta}
\right),
\end{equation}
where $N/V_H$ is the number of monopoles in the the present horizon 
volume, $V_H\equiv(4\pi/3)H_0^{-3}$, and $\Omega_0$ is the present 
density parameter. This 
indicates that the existence of even one monopole is critical for
$\eta\cong m_{\rm Pl,0}$. To see it more closely, we consider the condition that
the local density by one monopole does not exceed the cosmological 
mass density, {\it i.e.}, $\rho_{\rm gm}(R)<\rho_0$. This leads
\begin{equation}
R>{3H_0^{-1}\over\sqrt{\Omega_0}}\left({\eta\over m_{\rm Pl,0}}\right),
\end{equation}
which shows that the center of a monopole should be located farther than 
the horizon. A more stringent constraint is obtained by considering the 
difference of $\rho_{\rm gm}$ at both ends of the observed universe:
\begin{equation}
\Delta\rho_{\rm gm}={\eta^2\over(R-H_0^{-1})^2}-{\eta^2\over(R+H_0^{-1})^2}
\cong{4\eta^2H_0^{-1}\over R^3}.
\end{equation}
The condition $\Delta\rho_{\rm gm}/\rho_0<10^{-5}$ leads
\begin{equation}
R>{150H_0^{-1}\over\Omega_0^{\frac13}}
\left({\eta\over m_{\rm Pl,0}}\right)^{\frac23}.
\end{equation}

We now argue the plausible value of $R$ realized in this model. By 
assumption our universe was once in the core of a monopole with 
$\Psi\cong0$ and the Higgs field started classical evolution only 
after it had acquired an amplitude $\Psi
~\mbox{\raisebox{-1.0ex}{$\stackrel{\textstyle >}{\textstyle \sim}$ }}
\delta\Psi_{\rm Q}$. 
The number of $e$-folds after this epoch is approximately given by
\begin{equation}
N=2\pi\left({\eta\over m_{\rm Pl,0}}\right)^2
\ln\left(\sqrt{{6\pi\over\lambda}}{m_{\rm Pl,0}\over\eta}\right).
\end{equation}
For $\eta=m_{\rm Pl,0}$ and $\lambda=10^{-13}$, we find $N=103$. That is, 
the initial core radius becomes $e^{103-65}\sim10^{16}$ times as large 
as the present horizon. We therefore conclude that the core of a monopole 
is located so far from the observed region that even its long-range term 
$\rho_{\rm gm}=\eta^2/R^2$ does not exert effective astrophysical 
influence. Because this conclusion is irrelevant to the long-range term, 
it suggests that a local monopole in the same model is not detectable either.

\section{Summary and Discussion}

We have considered inflating monopoles in the BD theory. First, we 
have investigated the dynamics and spacetime structure of a global monopole. 
In the Einstein theory, monopoles undergo eternal inflation if and 
only if $\eta/m_{\rm Pl}
~\mbox{\raisebox{-1.0ex}{$\stackrel{\textstyle >}{\textstyle \sim}$ }}
0.33$. In the BD theory, on the other 
hand, the ratio $\eta/m_{\rm Pl}(\Phi)$ decreases until inflation ends. That is,
if $\eta/m_{\rm Pl}(\Phi)$ is large enough initially, a monopole starts to 
expand; however, after the ratio becomes smaller than the critical value 
$(\cong 0.33)$, the monopole turns to shrink. This behavior is quite different 
from that in the Einstein gravity and gives another mechanism to terminate 
inflation, which, however, is irrelevant to our universe.

Secondly, we have examined cosmological constraints on the 
parameters, which stems mainly from density perturbations. The COBE 
normalization requires 
$\eta/m_{\rm Pl,0}
~\mbox{\raisebox{-1.0ex}{$\stackrel{\textstyle >}{\textstyle \sim}$ }}
1$, $\lambda
~\mbox{\raisebox{-1.0ex}{$\stackrel{\textstyle <}{\textstyle \sim}$ }}
10^{-13}$ for $\omega>500$, implying that inflation in our universe must have 
ended in the standard manner as $\Psi$ reaches $\eta$. We 
also calculate the spectral index, which turns out to be typically 
around $n\cong 0.9$.

Although we have concentrated on the BD theory, finally we make a brief 
discussion on the behavior in general theories (\ref{action2}).
We still assume $U(\phi)=0$. 
The slow-roll conditions for the models are
\begin{equation}
{\beta^2\over2} ~ {\rm and} ~ |\beta\gamma|\ll 3,~~~
{e^{\gamma\kappa\phi/2}\over\kappa}\left|{V_{,\Psi}\over V}\right|
\ll\sqrt{6}~ {\rm and} ~
\sqrt{\left|{6\beta\over\gamma}\right|} ~ (\gamma\ne0),~~~
{e^{\gamma\kappa\phi}\over\kappa^2}\left|{V_{,\Psi\Psi}\over V}\right|\ll 3
\end{equation}
and the behavior of $\phi$ is given by
\begin{equation}
\phi-\phi_i={\beta\over\kappa}\ln{a\over a_i}.
\end{equation}
These show that, if $\beta\gamma>0$, $\gamma\phi$ increases and 
the slow-roll conditions eventually break down. We may therefore conclude 
that the finite duration of topological inflation is a general nature 
in generalized Einstein theories.

\acknowledgements

N. S. thanks Takashi Nakamura for discussions. We are grateful to 
Andrei Linde for useful comments. Numerical Computation of this work was 
carried out at the Yukawa Institute Computer Facility. N. S. was supported by 
JSPS Research Fellowships for Young Scientist. This work was supported partially 
by the Grant-in-Aid for Scientific Research Fund of the Ministry of Education, 
Science, Sports and Culture
(No.\ 9702603, No.\ 09740334, and Specially Promoted Research No.\ 08102010) 
and by the Waseda University Grant for Special Research Projects.

\newpage
\appendix
\section{Field Equations and Numerical Method}

In this Appendix we explain how we solve the field equations for a global 
monopole. As we shall show below, we improve our previous schemes \cite{SSTM}
so as to keep better accuracy.

The variation of (\ref{action}) with respect to $g_{\mu\nu}$, 
$\Phi^a$, and $\Phi$ yield the field equations:
\begin{eqnarray} \label{Ein}
G_{\mu\nu}\equiv{\cal R}_{\mu\nu}-\frac12g_{\mu\nu}{\cal R}
&=&{8\pi\over\Phi}T_{\mu\nu}+{\omega\over\Phi}
\left[\nabla_{\mu}\Phi\nabla_{\nu}\Phi
-{1\over2} g_{\mu\nu}(\nabla\Phi)^2\right]
+\nabla_{\mu}\nabla_{\nu}\Phi-g_{\mu\nu}\Box\Phi,\\
\label{Psi}
\Box\Psi^a&=&\frac{\partial V(\Psi)}{\partial\Psi^a},\\
\label{Phi}
\Box\Phi&=&\frac{8\pi T^{\sigma}_{\sigma}}{2\omega+3},
\end{eqnarray}
with
\begin{equation}
T_{\mu\nu}\equiv\nabla_{\mu}\Psi^a\nabla_{\nu}\Psi^a
-g_{\mu\nu}\left[\frac12(\nabla\Psi^a)^2+V(\Psi)\right].
\end{equation}

In the following we shall write down the field equations (\ref{Ein}), 
(\ref{Psi}), and (\ref{Phi}) with the metric (\ref{metric}) and the hedgehog 
ansatz (\ref{hg}). To begin with, we introduce the
extrinsic curvature tensor $K_{ij}$:
\begin{equation}
K^r_r=-{\dot A\over A},~~~ K^{\theta}_{\theta}(=K^{\varphi}_{\varphi})
=-{\dot B\over B},~~~
K\equiv K^i_i.
\end{equation}
Next, following Nakamura {\it et al.}\cite{Nak}, we define the following
variables to guarantee the regularity conditions at the center:
\begin{equation}
a\equiv{A-B\over r^2},~~~ k\equiv{K^{\theta}_{\theta}-K^r_r\over r^2},
\end{equation}
and introduce a new space variable, $x\equiv r^2$. Because only $\Psi$ is an odd
function of $r$ due to the hedgehog ansatz, we adopt a variable
$\psi\equiv\Psi/r$ instead of $\Psi$. 

There are several advantages in the above procedure of Nakamura {\it
et al.}\cite{Nak}~ The first is that any diverging factor at $r=0$ like 
$1/r$ does not appear in the basic equations below; hence we do not
have to treat the $r=0$ point separately. The second is that the new
variables $a$ and $k$ are replaced with differences between comparable 
quantities like $A-B$ at $r\approx 0$, which would generate numerical
errors due to a finite decimal of computers. The third is that a
derivative at $r\approx0$ is approximated by a finite difference much
more accurately. To see this, let us imagine a function
$F(r)$ which is expanded as
\begin{equation}
F(r)=c_0+c_1r+c_2r^2+c_3r^3+\cdots,
\end{equation}
around $r=0$. If we take the central difference of $F$ with respect to $r$, its
derivative at $r=\Delta r$ is approximated as
\begin{equation}
{dF\over dr}(\Delta r)={F(2\Delta r)-F(0)\over 2\Delta r}+O(\Delta r^2),
\end{equation}
that is, it is approximated only up to the $c_2$-term. If $F$ is given as
an even or odd function of $r$ and if we use a space variable $x$, on the other
hand, we can expand it as
\begin{eqnarray}
F_{{\rm even}}&=&c_0+c_2x+c_4x^2+\cdots,\\
{F_{{\rm odd}}\over r}&=&c_1+c_3x+c_5x^2+\cdots.
\end{eqnarray}
If we take the central difference with respect to $x$ at $x=(\Delta
r)^2$, for an even (odd) function, it approximates $dF/dx$ up to
$c_4~(c_5)$, and moreover guarantees $c_5=0~(c_6=0)$.

Furthermore, we define the following auxiliary variables to make the 
equations have a first-order form:
\begin{eqnarray}
&C\equiv-&{B'\over B}, \\
&\varpi\equiv\dot\psi, &~~~ \xi\equiv\psi',\\
&\Pi\equiv\dot\Phi, &~~~ \Xi\equiv\Phi',
\end{eqnarray}
where a prime and an overdot $\partial/\partial x$ and $\partial/\partial t$,
respectively. 
A full set of dynamical variables is 
$a,~B,~C$, $\psi,~\xi, ~\Phi,~\Xi$, $K,~k,~\varpi$, and $\Pi$. The 
time derivatives the first seven variables are given by the definitions above:
\begin{eqnarray}
&\dot a=-a K^r_r+&B k,\\
&\dot B=-B K^{\theta}_{\theta}, ~&~~~ \dot C={K^{\theta}_{\theta}}', 
\label{dotB}\\
&\dot\psi=\varpi, ~~&~~~ \dot\xi=\varpi',\\
&\dot\Phi=\Pi, ~~&~~~ \dot\Xi=\Pi'.
\end{eqnarray}
Note that the value of ${K^{\theta}_{\theta}}'$ in (\ref{dotB}) can be
determined by the momentum constraint (\ref{MC}), which gives a more accurate
value than a finite difference of $K^{\theta}_{\theta}$.

Now we write down the field equations (\ref{Ein}), (\ref{Psi}) and (\ref{Phi})
as
\begin{eqnarray}\label{HC}
-G^t_t&\equiv&{K^2-k^2x^2\over 3}
+{8x\over A^2}\left(C'-{A'C\over A}-{3C^2\over 2}\right)
+{4\over A^2}\left[{A'\over A}+4C+{a(A+B)\over 4B^2}\right] \nonumber\\
&=& {8\pi\over\Phi}\left[{x\varpi^2\over 2}+{2x\xi\over A^2}(x\xi+\psi)
+{\psi^2\over 2A^2}+{\psi^2\over B^2}+V \right] \nonumber\\
&& +{\Pi\over\Phi}\left({\omega\Pi\over 2\Phi}+K\right)
+{4\over A^2\Phi}\left[\Xi'+x\Xi
\left({\omega\Xi\over 2\Phi}-{A'\over A}-2C\right)+{3\Xi\over2} \right],
\\ \label{MC}
{G_{tr}\over 4r}&\equiv& {K^{\theta}_{\theta}}'+k\left({1\over2}-xC\right)
={2\pi\varpi\over\Phi}(2x\xi+\psi)
+{1\over 2\Phi}\left[\Pi'+\Xi({\omega\Pi\over\Phi}+K^r_{r})\right], \\
\label{dotK}
-{\cal R}^t_t&\equiv&\dot K-{K^2+2x^2k^2\over 3}
={8\pi\over\Phi} (x\varpi^2-V)-{12\pi T^{\sigma}_{\sigma}\over(2\omega+3)\Phi}
\nonumber\\ && ~~~~~~~~~~~~~~~~~~~~~
 +{\Pi\over\Phi}\left({\omega\Pi\over\Phi}+K\right)
+{4\over A^2\Phi}\left[x\Xi'-x\Xi\left
({A'\over A}+2C\right)+{3\Xi\over 2}\right],
\end{eqnarray}
\begin{eqnarray}\label{dotk}
&&{{\cal R}^r_r-{\cal R}^{\theta}_{\theta}\over r^2}\equiv
\dot k-kK+{2\over A^3}\left[2AC'-2A'C+aC+a'-{a^2\over B^2}
\left({A\over 2}+B\right)\right] \nonumber\\
&&~~={8\pi\over A^2\Phi}
\left[4\xi(x\xi+\psi)-{\psi^2 a(A+B)\over B^2}\right]
-{k\Pi\over\Phi}+{4\over A^2\Phi}\left[\Xi'
+\Xi\left({\omega\Xi\over\Phi}-{A'\over A}+C\right) \right],
\end{eqnarray}
\begin{eqnarray}\label{dotvarpi}
\dot\varpi&=&K\varpi+{4\over A^2}\left[x\xi'-x\xi\left({A'\over A}+2C\right)
+{5\xi\over 2} \right]
-{2\psi\over A^2}\left[{A'\over A}+2C+{a(A+B)\over B^2}\right] \nonumber\\
&&-\lambda\psi(\Psi^2-\eta^2),
\\ \label{dotPi}
\dot\Pi&=&K\Pi+{4\over A^2}\left[x\Xi'-x\Xi\left({A'\over A}+2C\right)
+{3\Xi\over 2}\right]-{8\pi T^{\sigma}_{\sigma}\over2\omega+3},
\end{eqnarray}
with
\begin{equation}
T^{\sigma}_{\sigma}\equiv x\varpi^2-{4x\xi\over A^2}(x\xi+\psi)
-\psi^2\left({1\over A^2}+{2\over B^2}\right)-4V.
\end{equation}

In order to set up initial data, we assume $A(t=0,r)=B(t=0,r)=1$ 
besides the matter configurations (\ref{Psi0}) and (\ref{Phi0}), and solve the
constraint equations (\ref{HC}) and (\ref{MC}) to determine $K(t=0,r)$ and
$k(t=0,r)$. Equation (\ref{dotK}), (\ref{dotk}), (\ref{dotvarpi}), and 
(\ref{dotPi}) provide the next time step of $K,~ k,~ \varpi$, and 
$\Pi$, respectively. The constraint equations
(\ref{HC}) and (\ref{MC}) remain unsolved during the evolution and are used for
checking the numerical accuracy. Through all the calculations the 
errors are always less than a few percent.


\baselineskip = 18pt

\begin{figure}
 \begin{center}
  \psbox[height=8cm]{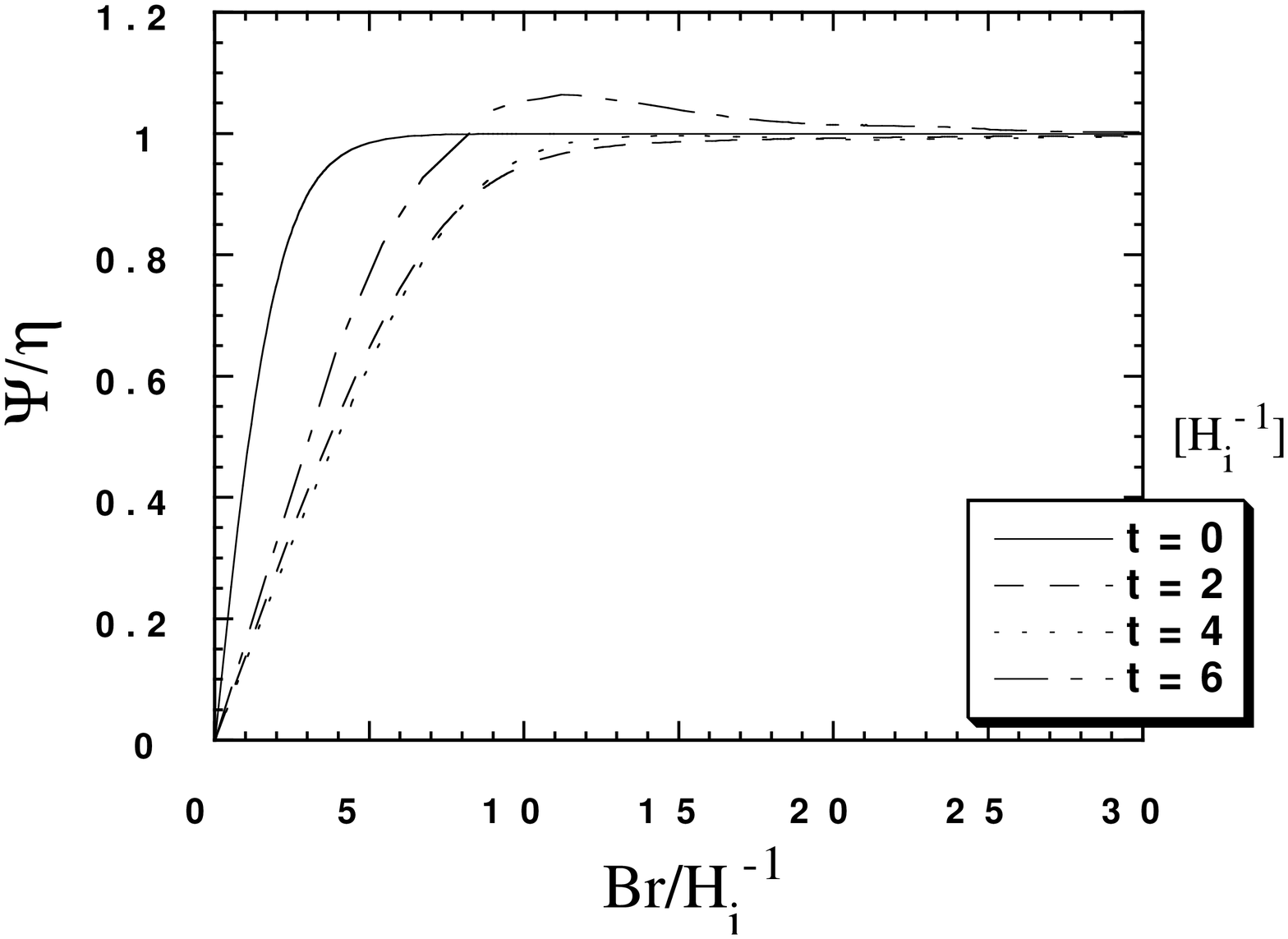}
  \psbox[height=8cm]{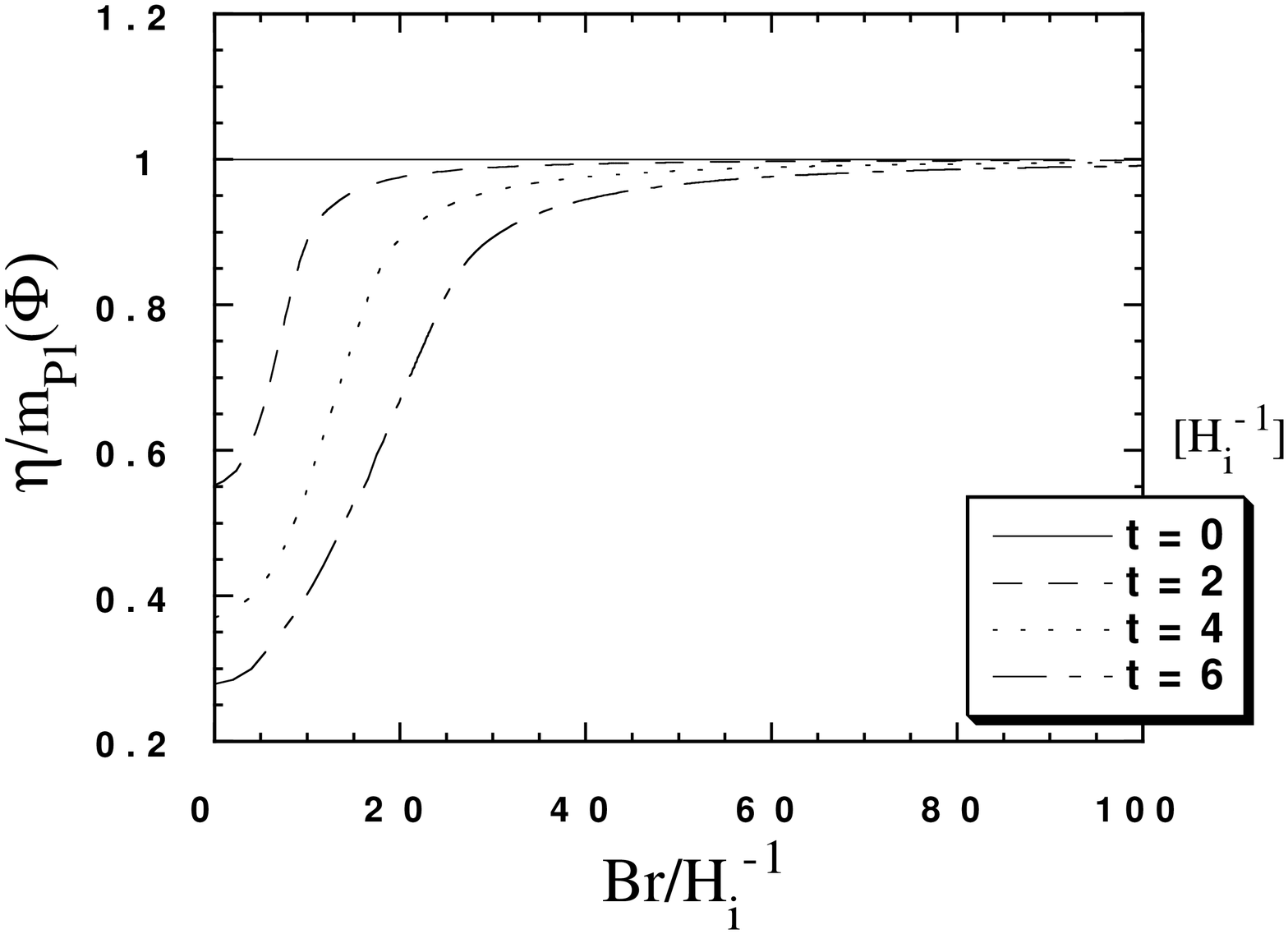}
 \end{center}
\end{figure}

\noindent
{\bf FIG.1}. Dynamics of a global monopole with $\omega=1$, $\lambda=0.1$, 
and $\eta/m_{\rm Pl}(\Phi_i)=1$. We plot $\Psi$ and $\eta/m_{\rm Pl}(\Phi)$ in 
(a) and in (b), respectively. 
Because we omit the far region where $\Psi$ is almost constant in (a), the 
scales of abscissas in (a) and in (b) are different.

\newpage
\begin{figure}
 \begin{center}
  \psbox[height=9cm]{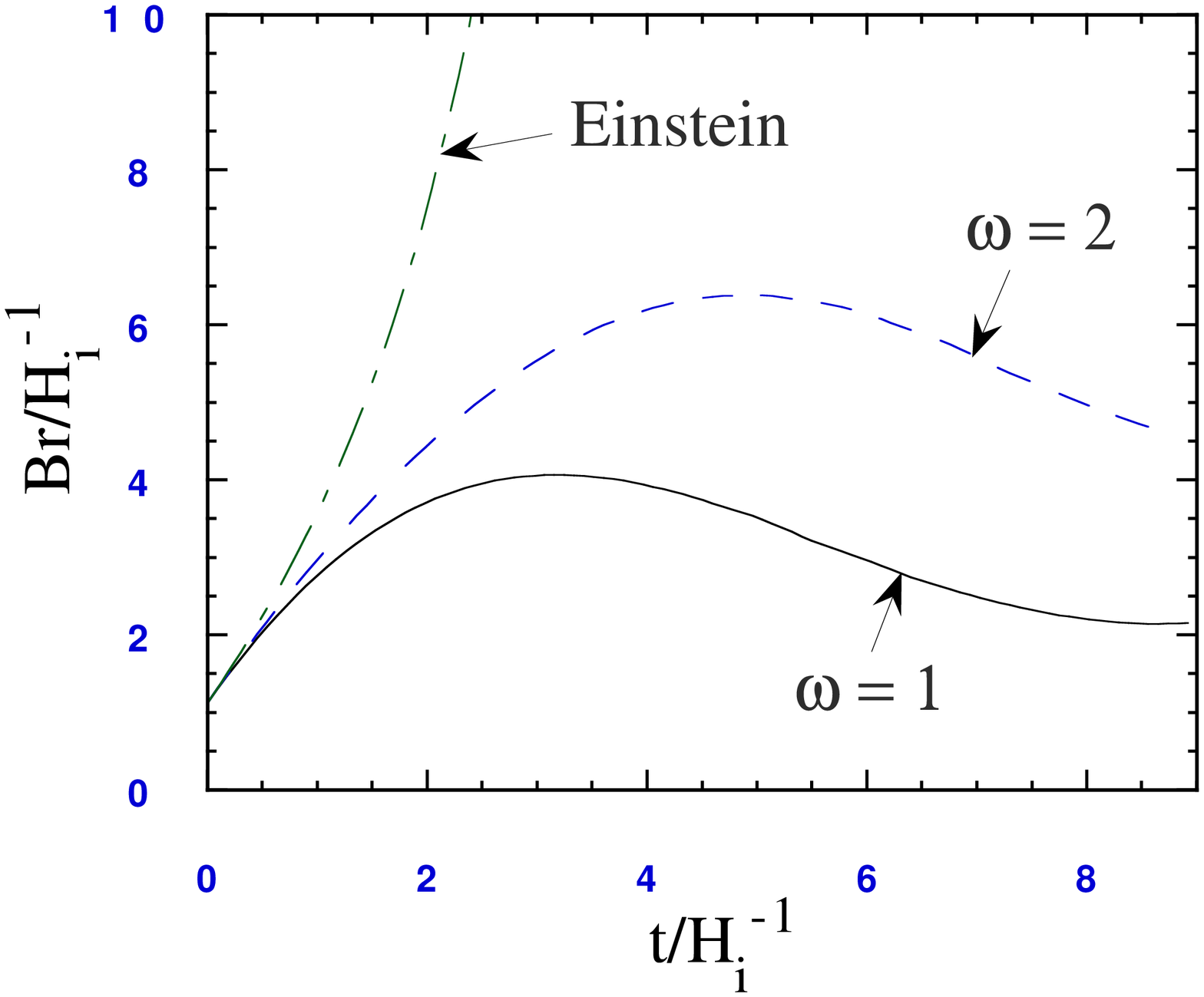}
 \end{center}
\end{figure}

\noindent
{\bf FIG.2}. Dependence of the evolution of a global monopole on 
$\omega$. We set $\lambda=0.1$ and $\eta/m_{\rm Pl}(\Phi_i)=1$. We plot
trajectories of the position of $\Psi=\eta/2$.

\newpage
\begin{figure}
 \begin{center}
  \psbox[height=9cm]{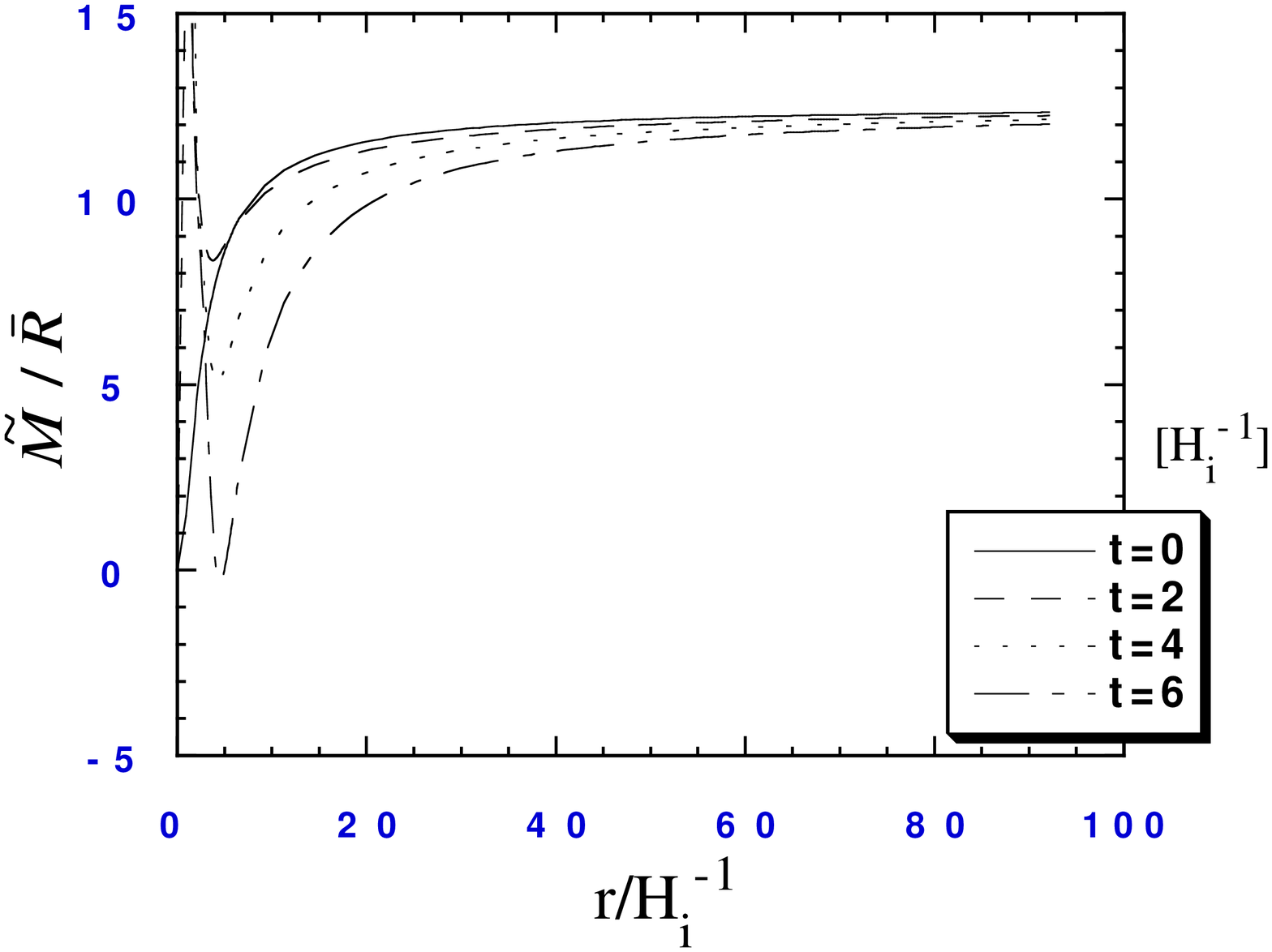}
  \psbox[height=9cm]{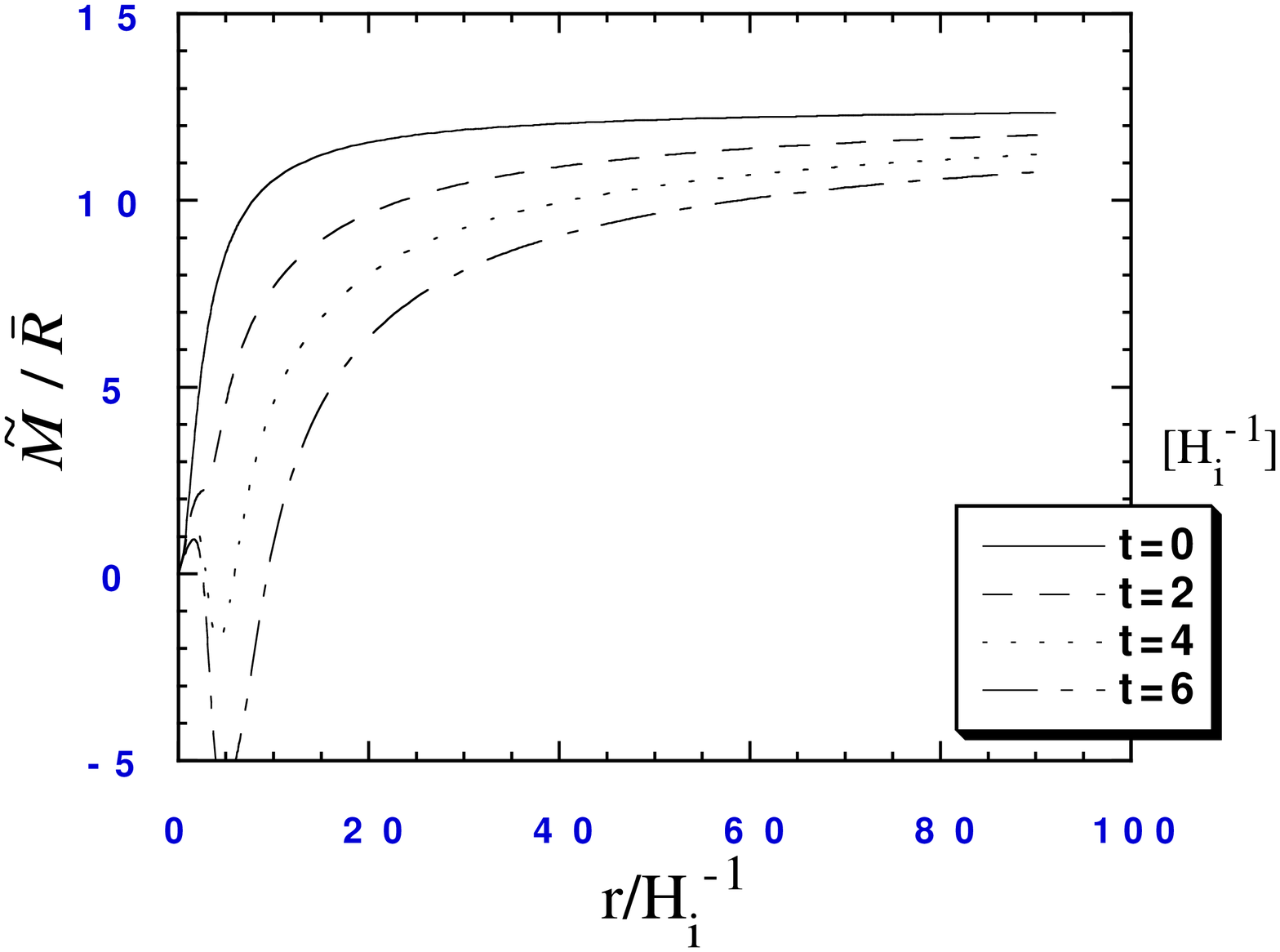}
 \end{center}
\end{figure}

\noindent
{\bf FIG. 3}. Behavior of the Misner-Sharp mass outside the core of a 
monopole. We take, for reference, the Einstein gravity in (a), and $\omega=1$
in (b). 

\newpage
\begin{figure}
 \begin{center}
  \psbox[height=8cm]{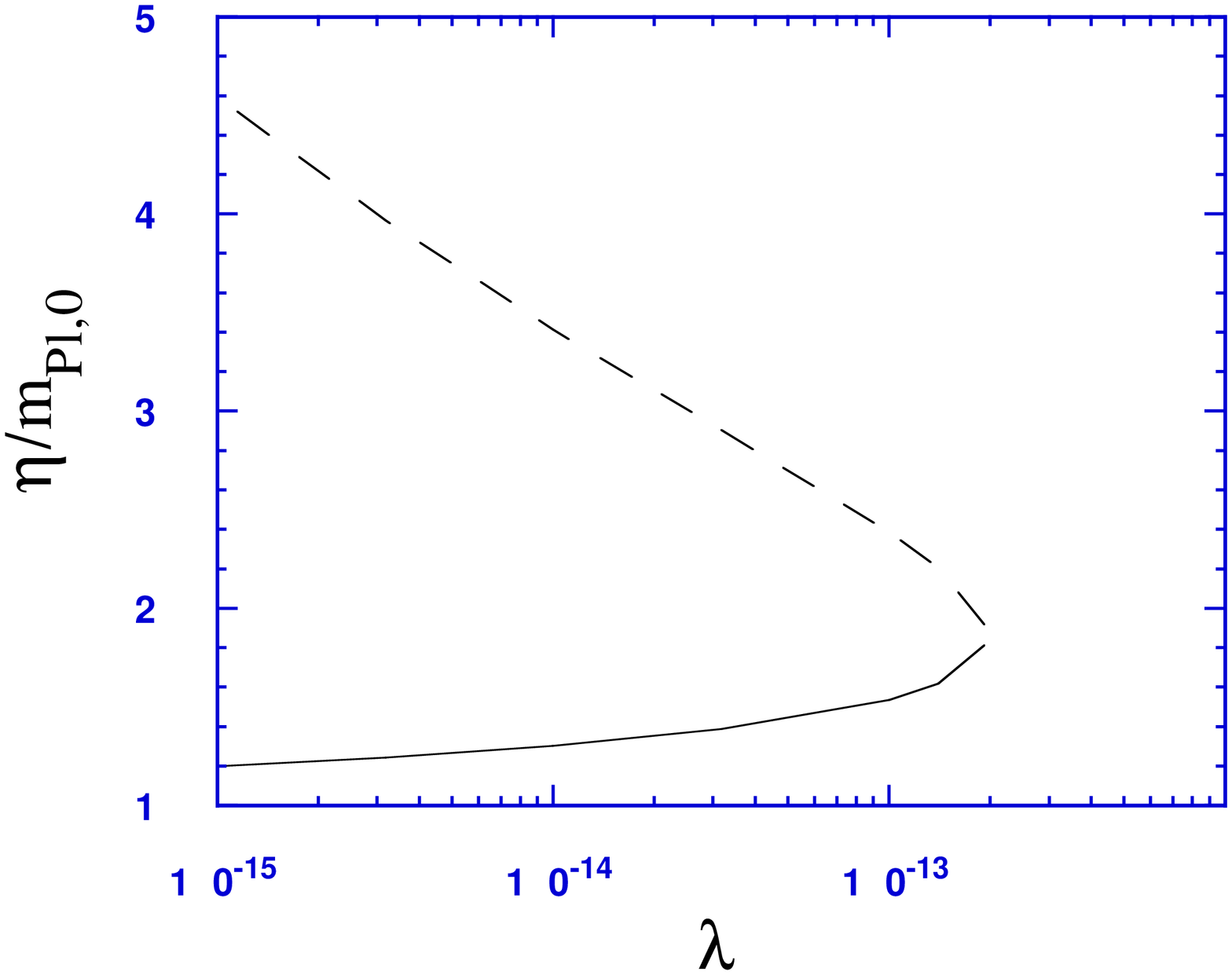}
 \end{center}
\end{figure}

\noindent
{\bf FIG. 4}. Concordant values of $\lambda$ and $\eta/m_{\rm Pl,0}$ with
COBE-normalized amplitudes of density perturbations.

\vskip .5cm
\begin{figure}
 \begin{center}
  \psbox[height=8cm]{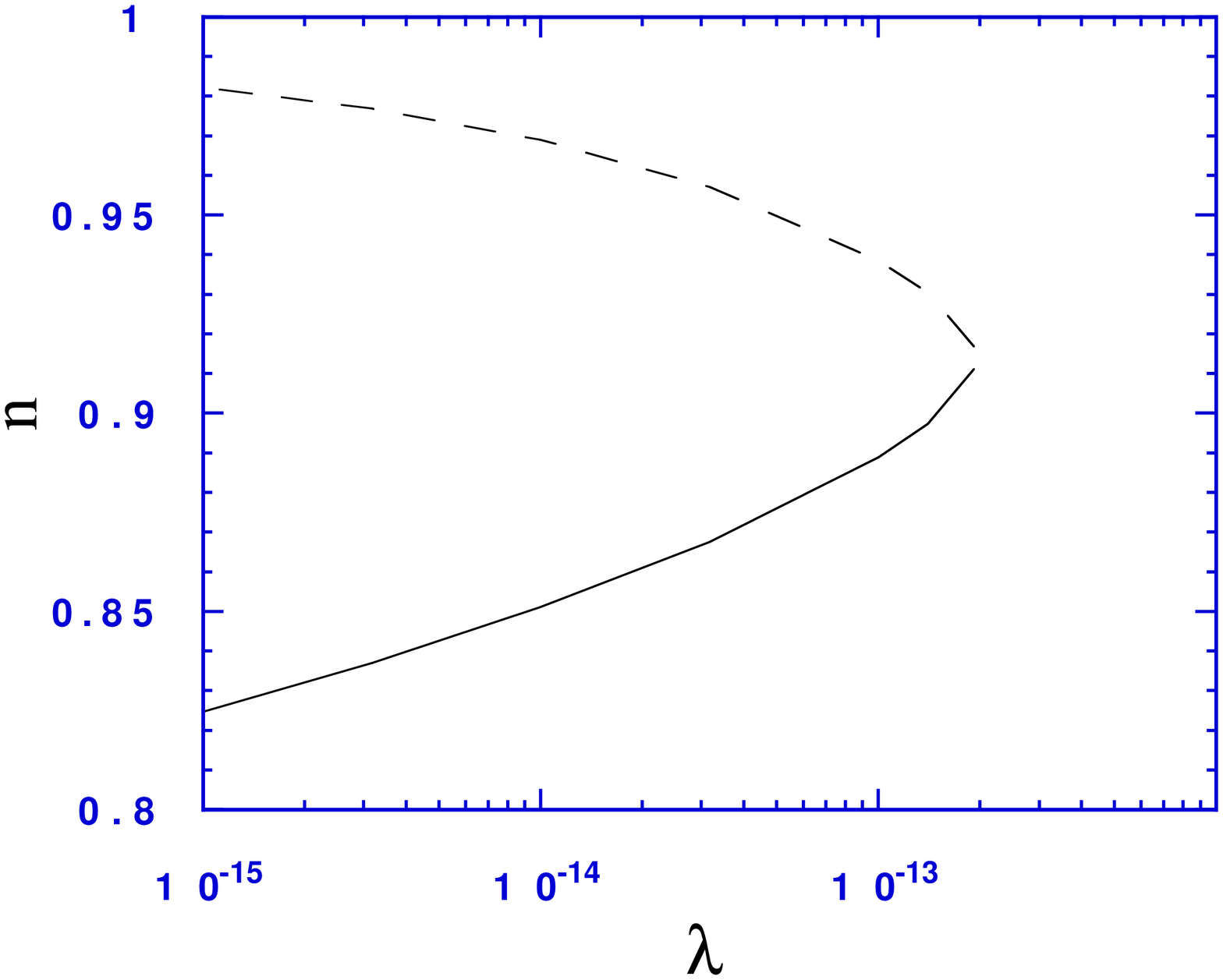}
 \end{center}
\end{figure}

\noindent
{\bf FIG. 5}. The spectral indices of density perturbations. These two 
index curves correspond to the two amplitude curves in Fig. 4.

\end{document}